# Feasibility Study for the Generation of High Power Continuous Wave Terahertz Radiation using Frequency Difference Generation

Author: Kathirvel Nallappan, PhD candidate in Engineering physics and Electrical Engineering, Polytechnique Montreal, Montréal, QC, Canada
Submitted for consideration to: Prof. R. Morandotti (INRS) and Prof. M. Skorobogatiy (Polytechnique)

**Abstract:** The goal of this report is to study the conditions for efficient THz generation using Frequency Difference Generation principle. In particular, two CW IR beams are used together with a periodically poled lithium niobate crystal in order to generate a continuous THz wave. As a source of the IR pump beams we consider a 30W (CW) IPG Er-doped fiber-amplifier which currently is the most powerful commercial amplifier on the market. The work is motivated by application of CW sources of THz waves in high bit wireless communication applications. The amplifier will allow to mix two optically modulated IR beams and to generate modulated THz signal.

## Calculation of the poling period:

In our studies, we have used a classic quasi-phase matching theory [1] that prescribes that efficient Difference frequency generation (DFG) in nonlinear media is achieved if both energy conservation and momentum conservation are achieved simultaneously by all the three waves involved in the process.

The following figure represents the wavenumbers of pump beams and difference wave and its phase mismatch that is typically observed in the dispersive media.

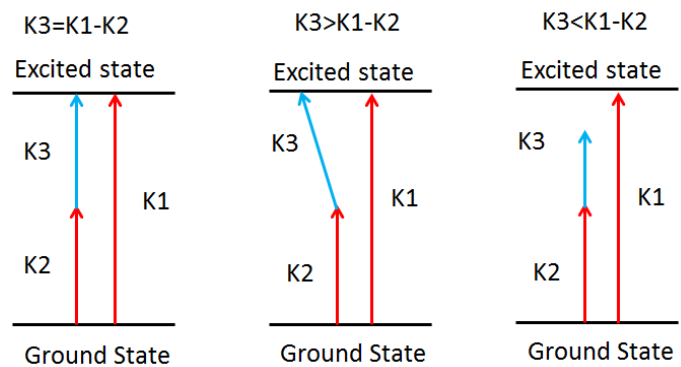

Fig 3: Wavevectors of the pump beams and the generated wave

Here K1 and K2 are the wavenumbers of the two IR pump beams; K3 is the wavenumber of the generated difference (THz) frequency.

In the case of non-dispersive media, the difference in the wavenumber of pump waves is always equal to the generated wave, i.e. K1=K2+K3. Or, equivalently:

$$K3 = K1 - K2$$

The wavenumber in a medium is calculated using the standard dispersion relation:

$$K = \frac{2\pi n(\lambda)}{\lambda}$$

However, due to material dispersion n1≠ n2 ≠ n3, one typically has a nonzero difference ΔK, thus preventing efficient difference frequency generation:

$$K3 + K2 - K1 = \Delta K \neq 0$$

In order to have growth in the generated wave, we have to compensate for the mismatch in the momenta of the three waves. This is commonly achieved by employing a Bragg grating written into the nonlinear media (poled lithium niobate crystal). According to the quasi-phase matching theory, the coherent length and the period of such a grating are given by:

$$\boldsymbol{coherent\ length = \frac{\pi}{\Delta K}}$$

$$\boldsymbol{poling\ period = 2 * coherent\ length}$$

As an example, in order to generate the difference frequency of 340 GHz, and for the pump beam wavelengths chosen as follows, we have:

$$\lambda 1 = 1.532\ \mu m\ (1.9568 * 10^{14} Hz)$$

$$\lambda 2 = 1.534666\ \mu m (1.95347 * 10^{14} Hz)$$

$$c/\lambda 1 - c/\lambda 2 = 340\ GHz$$

We have used the following Sellmeier equation for the lithium niobate crystal in the IR spectral range in order to calculate the refractive indices:

$$n^2 = 1 + \frac{2.6734\lambda^2}{\lambda^2 - 0.01764} + \frac{1.2290\lambda^2}{\lambda^2 - 0.05914} + \frac{12.614\lambda^2}{\lambda^2 - 474.60}$$

$n_1$(1.532 µm)= 2.2117
$n_2$(1.534666 µm)= 2.2116

For 340 GHz (881.74 µm), the refractive index $n_3$ of the LiNb is 5.2 from the literature [2-4]. Therefore, in the lithium niobate crystal we have:

$$K_1 = \frac{2\pi * 2.2117}{1.532 * 10^{-6}} = 9066237.598$$

$$K_2 = \frac{2\pi * 2.2116}{1.534666 * 10^{-6}} = 9050078.649$$

$$K_1 = \frac{2\pi * 5.2}{881.74 * 10^{-6}} = 37035.86091$$

$$K_3 + K_2 - K_1 = \Delta K = 20876.911 \text{m}$$

$$poling\ period = \frac{2\pi}{\Delta K} = 300.8 * 10^{-6} m$$

The value of **300 µm** corresponds well to the one found in the literature [2]. Similarly **for 260 GHz**, the calculation gives **397.28 µm** whereas in the literature [2] it is **396 µm**.

**Interaction length:**

The length of the crystal depends on the absorption coefficient for THz waves. The absorption coefficient of LiNb at 200 GHz is ~2 cm$^{-1}$ [2-3]. Thus, the effective interaction length in the absorptive medium, or in other words, the maximal useful length of a crystal, is given by:

$$L = \frac{20 * ln(2)}{a} = \frac{20 * ln(2)}{2\ cm^{-1}}$$

$$L = \sim 70\ mm$$

Which is again, matches well the results in the literature [2].

**Crystal Dimensions:**

In the literature, they typically use the crystal dimension of 0.5 mm X 0.5 mm. This height cannot be increased above 1 mm as the maximum thickness of the commercially available periodically poled lithium Niobate crystal is 1 mm.

The collimated beam diameter (1/e$^2$) from the 30W IPG laser amplifier is 1.1 mm, which in principle matches well with the maximal LiNb crystals available.

**Experimental Design:**

For the periodically poled crystal, we considered two system configurations, 1. Coupling of light into the crystal directly from the fiber, 2. Coupling of the collimated light (from free space) into the crystal.

*Coupling of light into the crystal directly from the fiber:*

If we launch the light from the SMF 28 fiber into the crystal, it will reflect back into the crystal as shown in figure 4, which can damage the amplifier.

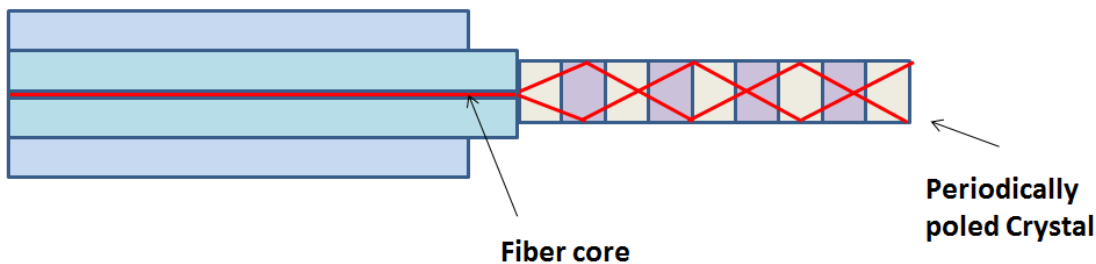

Fig 4: Butt coupling

*Coupling of the collimated light (from free space) into the crystal:*

In this approach, one can use telescopic arrangement to collimate the beam. In the EDFA amplifier specifications, the FWHM of the collimated beam is 1.1 mm. Therefore, at $1/e^2$ intensity, the beam waist can be calculated using the following relation.

$$W_0 = 0.8493218 * FWHM = 0.934 \, mm$$

As the laser amplifier is supplied with the integrated collimator, the following configuration is used to deliver the collimated beam with reduced beam size inside the crystal.

Using Thorlabs lenses with focal length of 18.40 mm and 2 mm respectively, the magnification factor would be 9.2 X. For the 0.934 mm beam, we will get the collimated beam of radius 0.101 mm.

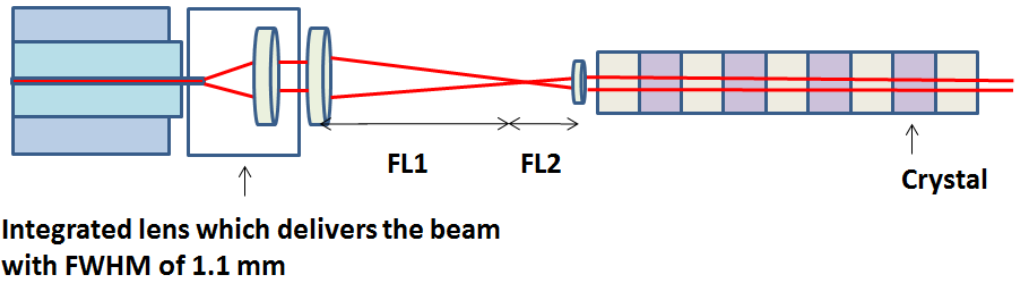

Integrated lens which delivers the beam
with FWHM of 1.1 mm

Fig 5: Delivery of the collimated pump beams into the crystal

**Theoretical simulation for DFG using crystal:**

The equation for DFG with quasi phase matching including losses is given [4].

$$\frac{P_3}{P_1} = \frac{8\pi^2 d_{eff}^2 L^2 P_2}{\varepsilon_0 c n_1 n_2 n_3 \lambda_3^2 A} * \exp(-(\alpha_1 + \alpha_2 + \alpha_3) * \left(\frac{L}{2}\right))$$

$$* \frac{Sin^2\left(\frac{\left(\Delta K - \frac{2\pi m}{\Lambda}\right)L}{2}\right) + \sinh^2\left[(\alpha_1 + \alpha_2 + \alpha_3)\left(\frac{L}{4}\right)\right]}{\left(\frac{\left(\Delta K - \frac{2\pi m}{\Lambda}\right)L}{2}\right)^2 + \left[(\alpha_1 + \alpha_2 + \alpha_3)\left(\frac{L}{4}\right)\right]^2}$$

Where,

$P_1$ & $P_2$=IR pump beams
$P_3$=Power of the generated wave (THz)
$d_{eff}$=(2/m$\pi$)*$d_{33}$. (Non-linear optical coefficients)
L=Length of the crystal
$\varepsilon_0$=Permittivity of free space
c=velocity of light
$n_1$ & $n_2$=Refractive index for the IR pump beams
$n_3$=Refractive index of the generated difference frequency (THz)
$\lambda_3$=Wavelength of the generated wave (THz)
A=Area of the beam
$\alpha_1$ & $\alpha_2$= Absorbance of IR pump beams
$\alpha_3$=Absorbance of generated wave (THz)
$\Delta K$=Difference in wave vector (K3+K2-K1)
m=Order of the grating
$\Lambda$=Period of the grating

The term $\frac{(\Delta K - \frac{2\pi}{\Lambda})L}{2}$ is equal to zero for the optimally chosen first order grating. Therefore, the equation above can be rewritten as

$$\frac{P_3}{P_1} = \frac{8\pi^2 d_{eff}^2 L^2 P_2}{\varepsilon_0 c n_1 n_2 n_3 \lambda_3^2 A} * \exp(-(\alpha_1 + \alpha_2 + \alpha_3) * \left(\frac{L}{2}\right)) * \frac{\sinh^2\left[(\alpha_1 + \alpha_2 + \alpha_3)\left(\frac{L}{4}\right)\right]}{\left[(\alpha_1 + \alpha_2 + \alpha_3)\left(\frac{L}{4}\right)\right]^2}$$

The following values are used in what follows: $d_{33}$=25 pm/V and $d_{eff}$=(2/π)$d_{33}$, r=100 microns and A=πr^2, $P_1$=$P_2$=15 Watts, $n_1$=2.1448, $n_2$=2.1447, $n_3$=5.1, $\alpha_1$=1 m$^{-1}$, $\alpha_2$=1 m$^{-1}$, $\alpha_3$=200 m$^{-1}$

The above equation is used for the generation of **140 GHz**. This wavelength is chosen due to commercial abundance of the THz detectors in this frequency range. It is difficult to find the exact value of the refractive index and absorption coefficient of the LiNb crystal below 200 GHz, therefore we use $n_3$=5.1, $\alpha_3$=200 m$^{-1}$ in our calculations. The crystal polling period is **764.20 μm for 140 GHz**. The simulated result shown below is for 30 W input power ($P_1$=$P_2$=15W as the total output power from the amplifier is 30 W, the power is divided equally among two wavelengths for the calculations) which is focused in a 100 μm beam radius.

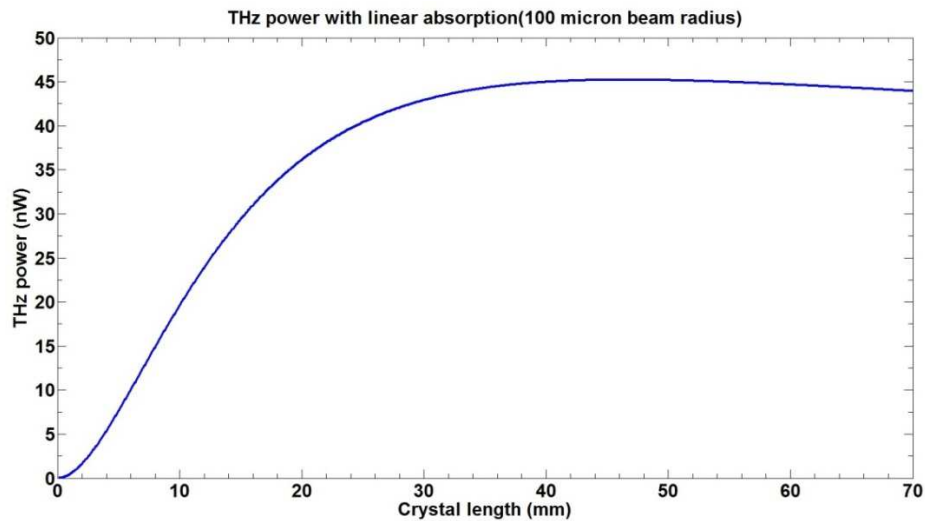

Fig 6: Simulation for generation of 140 GHz in a periodically poled lithium niobate using 30 Watt total CW input power ($P_1$=$P_2$=15W)

Main reason why the output power is low is because of the high value of the LiNb absorption coefficient $\alpha_3$=200 m$^{-1}$ at 140 GHz.

**Conclusion**

Theoretical analysis for the generation of THz frequency using DFG principle, IPG 30 Watt laser amplifier, and a LiNb periodically poled nonlinear crystal, predicts a relatively low output THz power of ~45 nW, which is mainly due to high LiNb absorption coefficient in THz. In order to improve on this result, one has to search for the higher nonlinearlity, and lower absorption nonlinear media.